\documentstyle[psfig]{mn}

\newif\ifAMStwofonts
\AMStwofontstrue

\ifoldfss

  \newcommand{\bld}[1] {{\bf #1}}
  \ifCUPmtlplainloaded \else
    \NewTextAlphabet{textbfit} {cmbxti10} {}
    \NewTextAlphabet{textbfss} {cmssbx10} {}
    \NewMathAlphabet{mathbfit} {cmbxti10} {} 
    \NewMathAlphabet{mathbfss} {cmssbx10} {} 
  \fi
  \ifAMStwofonts
    \ifCUPmtlplainloaded \else
      \NewSymbolFont{upmath} {eurm10}
      \NewSymbolFont{AMSa} {msam10}
      \NewMathSymbol{\upi}     {0}{upmath}{19}
      \NewMathSymbol{\umu}     {0}{upmath}{16}
      \NewMathSymbol{\upartial}{0}{upmath}{40}
      \NewMathSymbol{\leqslant}{3}{AMSa}{36}
      \NewMathSymbol{\geqslant}{3}{AMSa}{3E}
      \let\oldle=\le

    \fi
  \fi
\fi 

\ifnfssone
  \newmathalphabet{\mathit}
  \addtoversion{normal}{\mathit}{cmr}{m}{it}
  \addtoversion{bold}{\mathit}{cmr}{bx}{it}

  \newcommand{\bld}[1] {\mathbf{#1}}
  \newmathalphabet{\mathbfit} 
  \addtoversion{normal}{\mathbfit}{cmr}{bx}{it}
  \addtoversion{bold}{\mathbfit}{cmr}{bx}{it}
  \newmathalphabet{\mathbfss} 
  \addtoversion{normal}{\mathbfss}{cmss}{bx}{n}
  \addtoversion{bold}{\mathbfss}{cmss}{bx}{n}
  \ifAMStwofonts
    \ifCUPmtlplainloaded \else
      \UseAMStwoboldmath
      \makeatletter
      \new@mathgroup\upmath@group
      \define@mathgroup\mv@normal\upmath@group{eur}{m}{n}
      \define@mathgroup\mv@bold\upmath@group{eur}{b}{n}
      \edef\UPM{\hexnumber\upmath@group}
      \new@mathgroup\amsa@group
      \define@mathgroup\mv@normal\amsa@group{msa}{m}{n}
      \define@mathgroup\mv@bold\amsa@group{msa}{m}{n}
      \edef\AMSa{\hexnumber\amsa@group}
      \makeatother
      \mathchardef\upi="0\UPM19
      \mathchardef\umu="0\UPM16
      \mathchardef\upartial="0\UPM40
       \mathchardef\leqslant="3\AMSa36
      \mathchardef\geqslant="3\AMSa3E
      \let\oldle=\le

    \fi
  \fi
\fi 

\ifnfsstwo

  \newcommand{\bld}[1] {\mathbf{#1}}
  \DeclareMathAlphabet{\mathbfit}{OT1}{cmr}{bx}{it}
  \SetMathAlphabet\mathbfit{bold}{OT1}{cmr}{bx}{it}
  \DeclareMathAlphabet{\mathbfss}{OT1}{cmss}{bx}{n}
  \SetMathAlphabet\mathbfss{bold}{OT1}{cmss}{bx}{n}
  \ifAMStwofonts
    \ifCUPmtlplainloaded \else
      \DeclareSymbolFont{UPM}{U}{eur}{m}{n}
      \SetSymbolFont{UPM}{bold}{U}{eur}{b}{n}
      \DeclareSymbolFont{AMSa}{U}{msa}{m}{n}
      \DeclareMathSymbol{\upi}{0}{UPM}{"19}
      \DeclareMathSymbol{\umu}{0}{UPM}{"16}
      \DeclareMathSymbol{\upartial}{0}{UPM}{"40}
      \DeclareMathSymbol{\leqslant}{3}{AMSa}{"36}
      \DeclareMathSymbol{\geqslant}{3}{AMSa}{"3E}
      \let\oldle=\le

    \fi
  \fi
\fi 

\ifCUPmtlplainloaded \else
  \ifAMStwofonts \else 
    \def\upi{\pi}
    \def\umu{\mu}
    \def\upartial{\partial}
  \fi
\fi

\title{Principal Component Analysis of Synthetic Galaxy Spectra}
\author[Shai Ronen, Alfonso Arag\'on-Salamanca, Ofer Lahav]
       {Shai Ronen, Alfonso Arag\'on-Salamanca, Ofer Lahav \\
        Institute of Astronomy, The Observatories, Madingley Road, Cambridge, CB3 OHA}

\date{Submitted to MNRAS, 8 May 1998 }

\pagerange{\pageref{firstpage}--\pageref{lastpage}}
\pubyear{1998}

\begin{document}

\maketitle

\label{firstpage}

\begin{abstract}

We analyse synthetic galaxy spectra from the evolutionary models of
Bruzual \& Charlot and Fioc \& Rocca-Volmerange using the method of
Principal Component Analysis (PCA).  We explore synthetic spectra with
different ages, star formation histories and metalicities, and
identify the Principal Components (PCs) of variance in the spectra due
to these different model parameters.  The PCA provides a more objective
and informative alternative to diagnostics by individual spectral
lines.  We discuss how the PCs can be used to estimate the input model
parameters and explore the impact of noise in this inverse problem.
We also discuss how changing the sampling of the ages and other model
parameters affects the resulting PCs. Our first two synthetic PCs
agree with a similar analysis on observed spectra obtained by
Kennicutt and the 2dF redshift survey.  We conclude that with a good
enough signal-to-noise ($S/N\gg 10$) it is possible to derive age, star
formation history and metallicity from observed galaxy spectra using
PCA.

\end{abstract}


\section{Introduction}

Present and future galaxy surveys will provide large numbers of
spectra, which will allow the determination of many properties of
different galaxy populations.  For example, the Anglo-Australian
Observatory 2-degree-Field (2dF) Galaxy Survey aims to collect 250,000
spectra and is already gathering data (e.g. Colless 1998). 
The Sloan Digital Sky Survey (e.g. Gunn \& Weinberg 1995; Fukugita 1998)
will observe the spectra of millions of galaxies.
Such large data sets
will provide a wealth of information, pertaining to the distribution
and properties for a vast variety of galaxy types.  
The analysis and full exploitation
of such data sets will require well understood, automated and objective
techniques to extract as much information as possible on the
evolutionary properties of the galaxies. In this paper we
present the  method of Principal Component Analysis (PCA) 
to extract such information by looking at
the underlying parameters that determine each galaxy spectrum. We apply
the method to synthetic galaxy spectra generated using the evolutionary
population synthesis technique (see, e.g.,  Bruzual \& Charlot 1993;
Fioc \& Rocca-Volmerange 1997). For such model spectra, the evolutionary
properties are known {\it a priori\/}, allowing us to relate the 
information provided by the PCA to the input model parameters. 

Three of the most important parameters which determine a galaxy
spectrum are age, star formation history and metallicity. Given a
spectrum, one would like to deduce these parameters. This can be done
in principle by comparing the given spectrum with synthetic spectra
based on models of stellar evolution.  Attempts have been made in this
direction using template fitting. Summarising these attempts, Nobuo
Arimoto (in Leitherer et al. 1996) writes that `It turns out that the
present-day population synthesis models cannot derive age and
metallicity of stellar populations, independently, from the
photo-spectroscopic features of elliptical galaxies'. We note, however,
that improvement in the statistical analysis technique may also help,
and methods more sophisticated than simple template fitting are
available.

In this paper we explore a standard statistical technique, Principal
Component Analysis, as a promising approach which may prove better than
template fitting. PCA has previously been applied to data
compressing  and classification of spectral data 
of stars (e.g. Murtagh \& Heck 1987; Bailer-Jones et al. 1997), 
QSO (e.g. Francis et al. 1992) and galaxies 
(e.g. Connoly et al. 1995; 
Folkes, Lahav \& Maddox  1996;
Sodre \& Cuevas 1997;
Galaz \& de Lapparent, 1997; Bromley et al. 1998;  
Glazebrook, Offer \& Deeley 1998).  
Here we apply it to samples of synthetic spectra in the
$3650$--$7100$\AA\ rest-frame wavelength range (similar
to that of the 2dF and Sloan surveys) to get physical insight into the
way different parameters affect the spectrum of a galaxy. We leave the
comparison with observations to a future paper. However we already
notice an encouraging good resemblance between the PCs that we  get
from the synthetic spectra and those one gets from Kennicutt's (1992)
data of field galaxies.

The organisation of the rest of this paper is as follows. In section~2
we describe the synthetic models we have used. In section~3 we present
the PCA method. In section~4 we analyse a sample of instantaneous burst
galaxy models with different ages, find the first Principal Components
(PCs), discuss their physical significance, their use to determine the
age, and how different age sampling affects the resulting PCs. In
section~5 we explore galaxy models with different ages and different
star formation histories, and discuss how PCA may enable us to
discriminate between the effect of these two parameters on the spectra.
In section~6 we quantify the effect of varying metallicity on the
spectra.  In section~7,  we check how noise in the data may affect the
discriminatory power of our method. We end with a discussion and the
conclusions of our investigation.

\section[]{The Synthetic Models}

We have chosen to use two different synthetic models from different
authors so that we can assess how model-dependent our conclusions are.
The first one is PEGASE (Projet d'Etude des GAlaxies par Synth\`ese
Evolutive), by Fioc \& Rocca-Volmerange (1997). The second one is the
isochrone synthesis model of Bruzual \& Charlot (1996 ---GISSEL: Galaxy
Isochrone Synthesis Spectral Evolution Library).  Given a Star
Formation Rate history, an Initial Mass Function and metallicity (only
solar metallicity for PEGASE), these models provide galaxy spectra as a
function of age.  We used these models to simulate spectra in the
optical range $\lambda\lambda3650$--$7100$\AA, appropriate for the 2dF
and Sloan surveys.

The PEGASE model extends from the UV to the NIR and is based on a
combination of observational and synthetic libraries of stellar
spectra. The optical range in which we are interested is based on the
library of Gunn \& Stryker (1983), and has $10$\AA\ spectral bins.  One
advantage of this model is that, on top of including the contribution
to the spectra of the stellar population, it also calculates the
intensities of Nebular emission lines.  Since the shape of the emission
lines is not specified, for the purpose of this paper we simulated the
shape of all emission lines as a Gaussian with a sigma of 10 angstroms,
i.e., implying a spectral resolution comparable to that of the stellar
component in the models. At such resolution, the emission lines will be
unresolved for normal galaxy spectra (i.e., not AGNs), thus the shape
and width of the lines would be determined by the instrumental
resolution.  The origin of the nebular emission lines is the absorption
of Lyman continuum photons below $912$\AA\ by the gas, which gets
ionised, and reaches recombination equilibrium (cf. Osterbrock 1989).
We let the fraction of absorbed photons to be 0.7 in all our
simulations \footnote{In the synthetic spectra, the
[OIII] $\lambda4959$\AA--$\lambda5007$\AA\ doublet is represented by the
$\lambda5007$\AA\ line only. Since the intensity of the
$\lambda4959$\AA\ line is always $1\over3$ that of the
$\lambda5007$\AA\ line, there is no loss of information.}.

The Bruzual \& Charlot (1996; BC) isochrone model of stellar population
synthesis is widely used in the interpretation of observed galaxy
spectra.  In the optical range, and for solar metallicity stellar
populations, it offers a choice between three stellar libraries. We
have confirmed that both PEGASE and BC models give very similar spectra
when both are used to simulate populations of different ages based on
the same stellar library (e.g., Gunn \& Stryker 1983), and with no emission
lines. The disadvantage of the BC model is that it does not provide
emission lines. However the last version of the BC models have the
advantage of providing spectra for different metalicities. These are
based on the theoretical stellar spectra compiled by Lejeune, Cuisinier 
 \& Buser (1996).  We therefore chose this model for exploring the
impact of different metalicities on the spectra using PCA.

In both models we used the Initial Mass Function (IMF) 
of Scalo (1986), which is representative of the solar neighbourhood.

Note that we have not included the effects of interstellar reddening
in the models for simplicity, since the main aim of this paper is
to illustrate the PCA technique. It is important to bear in mind
that reddening would add another important parameter affecting the
galaxy spectra, in particular the shape of the continuum and the relative
intensities of the emission lines, although not the equivalent width
of the absorption features. 

\section[]{Principal Component Analysis}

Suppose we have a sample of $N$ galaxy spectra, all covering the same
rest-frame wavelength range. Each spectrum is described by an
$M$-dimensional vector $X$ containing the galaxy flux at $M$ uniformly
sampled wavelengths. Let $S$ be the $M$-dimensional space spanned by
the `spectral' vectors $X$. A given spectrum, then, is a point in
$S$-space, and the spectra in the sample form a cloud of points in
$S$.  The position of a galaxy spectrum point in this space depends on
parameters such as age, star formation history and metallicity. However
it is impossible to visualise directly how the data are distributed in
high dimensional spaces. An alternative is to employ some technique for
dimension reduction by projecting the data in, say, two dimensions.

Here we analyse the data with a standard technique, Principal Component
Analysis (PCA).  commonly used in Astronomy (e.g, Murtagh \& Heck
(1987) and references therein).  The PCA method is also known in the
literature as Karhunen-Lo\'eve or Hotelling transform.

PCA is an orthogonal transformation that allows the building of more
compact linear combinations of data that are optimal with respect to
the mean square error criterion. The new orthogonal basis is composed
of vectors called Principal Components (PCs). The first Principal
Component is taken to be along the direction in the $S$ space with the
maximum variance of data points $X$. More generally, the $k$-th
component is taken along the maximum variance direction in the subspace
perpendicular to the subspace spanned by the first $(k-1)$ Principal
Components.

The formulation of standard PCA is as follows (see
 e.g. Murtagh \& Heck 1987). Consider a set of $N$
objects ($i=1,N$), each with $M$ parameters ($j=1,M$). If $r_{ij}$ are
the original measurements, we construct normalised properties as
follows:
\begin{equation}
X_{ij}=r_{ij}-\overline{r}_j,
\end{equation}
where
\begin{equation}
\overline{r}_j={1\over N}\sum_{i=1}^N r_{ij}
\end{equation}
is the mean. We then construct a covariance matrix
\begin{equation}
C_{jk}={1\over N}\sum_{i=1}^N X_{ij}X_{jk} \quad 1\oldle j\oldle M \quad 1\oldle k\oldle M. 
\end{equation}
It can be shown that the axis along which the variance is maximal is the eigenvector $\bmath{e_1}$ of the matrix equation
\begin{equation}
\bld{C}\bmath{e_1}=\lambda_1\bmath{e_1,}
\end{equation}
where $\lambda_1$ is the largest eigenvalue, which is in fact the
variance along the new axis. The other principal axes and eigenvectors
obey similar equations. It is convenient to sort them in decreasing
order, and to quantify the fractional variance by
$\lambda_\alpha/\sum_\alpha\lambda_\alpha$. It is also convenient to
renormalize each component by $\sqrt\lambda_\alpha$, to give unit
variance along the new axis. 
The sign ($+/-$)  of the eigen-vectors is arbitrary. 
We note that the weakness of PCA is that
it assumes linearity and also depends on the way the variables are
scaled. In the analysis which follows, the spectra are all normalised
to have the same total flux over the wavelength range considered.  This
is a simple and convenient choice which preserves the shape of the
spectrum.

The matrix of all the eigenvectors forms a new set of orthogonal axes
which are ideally suited to a description of the data set.  If the main
variance of the data set lies in a small dimensional space, then one
can get a good visualisation of it by plotting the data projected on
the first few eigenvectors (PCs), the projection of a vector $v$ on the
eigenvector $\bmath{e_i}$ being $\bmath{v\cdot e_i}$. The fractional
variance of the first PCs tells us to what extent the data lie in a
given low dimensional space. In what follows, we will apply this
technique to different model galaxy samples.
To avoid confusion we denote eigenvectors by PC1,PC2 etc.
and projections on the new axes by pc1, pc2 etc.
\section[]{Single burst galaxies}

\begin{figure}
 \psfig{figure=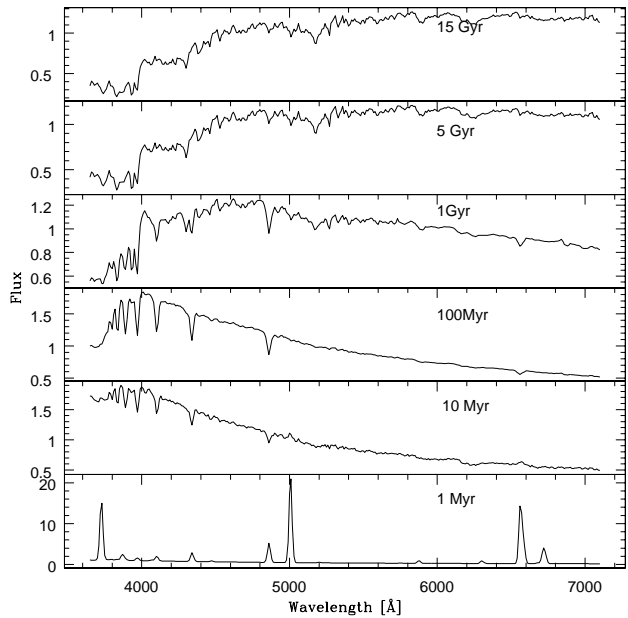,width=3.5in} 
 \caption{ {\bf Model spectra} for different galaxy ages computed with the
PEGASE code for stellar populations formed in an instantaneous burst.
The spectra are normalised to the same constant total flux over the
wavelength range of interest.}
\end{figure}

We used PEGASE to simulate the spectra of galaxies formed in a single
instantaneous burst that takes place at $t=0$ in the age range
0--18.5$\,$Gyr.  Figure~1 shows some of the model spectra at different
ages. For our first study, we simulated 1850 spectra between 10$\,$Myr
and 18.5$\,$Gyr in equal time intervals of 10 Myr, and performed PCA on
this sample. However, we found that by 10 Myr emission lines already
decay to a negligible value compared with the continuum. To see the
change in emission lines in very early ages, we also performed  PCA on
a sample of 15 spectra with ages 0--15$\,$Myr and time step of
1$\,$Myr.

Figure~2 shows the mean of the 1850 spectra along with the first three
eigenvectors. The first PC (PC1) accounts for 98.5\% of the variation
in the spectra. It correlates blue continuum with Balmer absorption
lines.  What seems like some emission in the Mgb and NaD lines
represents compensation for the absorption in this lines in the average
spectrum.  Also note that the blue spectrum is correlated with the
negative of broad absorption features in the red end (6000--7000\AA)
which appear in the average spectrum. We identify these as typical
molecular absorption lines of an M-star spectrum. The physical
significance of this is that as a galaxy ages, later type stars
contribute more relative flux to the integrated spectrum and it becomes
redder, so the M-star spectral features should be anti-correlated with
blue colour.  The second principal component, PC2, accounts for only
0.9\% of the variation, therefore any features present in it (e.g., the
strong Ca K absorption) represent only fine tuning to the main
variation which PC1 represent.  PC3 accounts for only 0.5\% of the
variation. It correlates the Balmer absorption lines with absorption
below 4000\AA, i.e, the Balmer break.

To demonstrate the use of the PCA to determine the age of a galaxy from
its spectrum we plot in Figure~3 the projection on the first PC (pc1)
versus age. This is mainly a monotonically decreasing curve, clearly
showing the reddening of the galaxy spectrum with age. Note that the
slope becomes much shallower at later ages, reflecting the slower
change in spectral shape at old ages since the lifetime of the main
stellar contributors becomes larger.  This makes accurate age
determinations more difficult for old galaxies. Some small
non-monotonic irregularities in the curve are probably due to abrupt
transitions  of stars to different evolutionary stages. They may be
real, but they may also occur if the model does not contain enough
evolutionary tracks in a certain range. In any case, we shall see that
these irregularities will be smoothed out when considering longer and
more realistic (continuous) star formation histories instead of the
instantaneous bursts considered here.

In figure~4 we plot the projection on the PC1--PC2 plane. The turning
point in the graph correspond to an age around 400$\,$Myr, and the PC2
maximum corresponds to 770$\,$Myr.  Since one of the the dominant
features of PC2 is the Ca~K absorption line, this suggests a maximum
for K absorption around that time. To check this hypotheses, we plot in 
figure~5 the equivalent width of the K absorption line versus age.
There is a sharp rise in absorption at early ages, and a turning at 
2$\,$Gyrs to a rather flat curve.  Overall there is a good correspondence
with the PC2 behaviour, but the maximum occurs at later age. This
happens because PC2 contains other features as well, in particular a
strong `Balmer decrement' below $\simeq4000$\AA\ which makes the
projection negative for very blue young galaxies, and a slightly blue
slope above 4000\AA\ that makes the projection become negative for old
galaxies. Note that the Ca~H line appears to behave  differently from
the Ca~K line. This is due to the close proximity of the Balmer
H$\epsilon$ line, whose strength often anti-correlates with that of 
the Ca~H line.

The Ca~K line and other metallic absorptions appear in a different PC
than the Balmer lines because the metal lines develop later than the
Balmer series, as an examination of figure~1 reveals.  This is easily
understood in terms of the spectral properties of the dominant stellar
types at each age: young A spectral type stars dominate when the galaxy
is $\simeq1\,$Gyr old, while stars with strong metallic lines have
lower masses, thus dominating at later ages.  The hydrogen absorption
lines are correlated with the blue spectrum of PC1, while the metallic
absorptions are correlated with the redder spectrum of PC2 and with the
4000\AA\ break. These are typical features of a spectrum older than
1$\,$Gyr (figure~1), where the metallic absorptions (e.g., Ca~K) are
strong. This simple exercise demonstrates how PCA, which does not
`know' anything about stellar evolution or stellar spectra is able to
identify the main contributors to a galaxy spectrum as it evolves.

\begin{figure}
 \psfig{figure=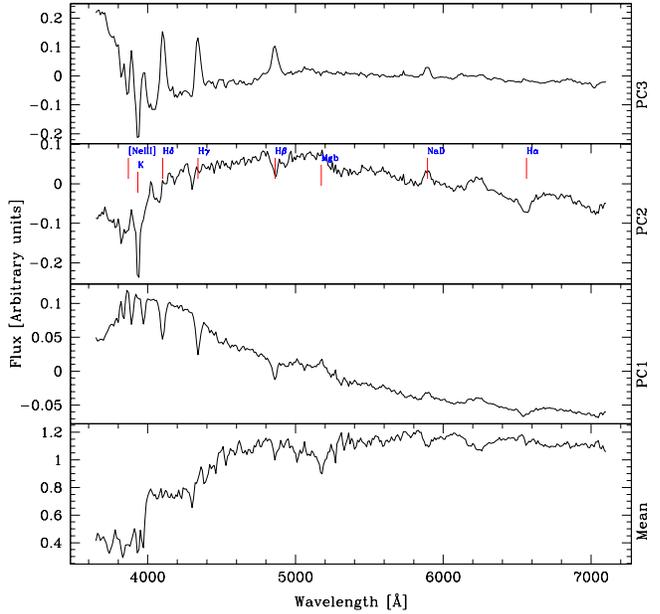,width=3.5in} 
 \caption{ {\bf Single burst:} the mean of the 1850 synthetic spectra of galaxies formed in
a single instantaneous burst that takes place at $t=0$ in the age range
0.01--18.5$\,$Gyr in equal time intervals of 10$\,$Myr (bottom).  The
first three PCs are plotted above it.  The wavelength of some atomic
line features are marked. Note that these may represent either emission
or absorption, depending on the sign of the weight factor multiplying
each PC. The PC1,PC2 and PC3 correspond to a variance
in the spectra of 98.5\%, 0.9\% and 0.5\%.
}

\end{figure}
\begin{figure}
 \psfig{figure=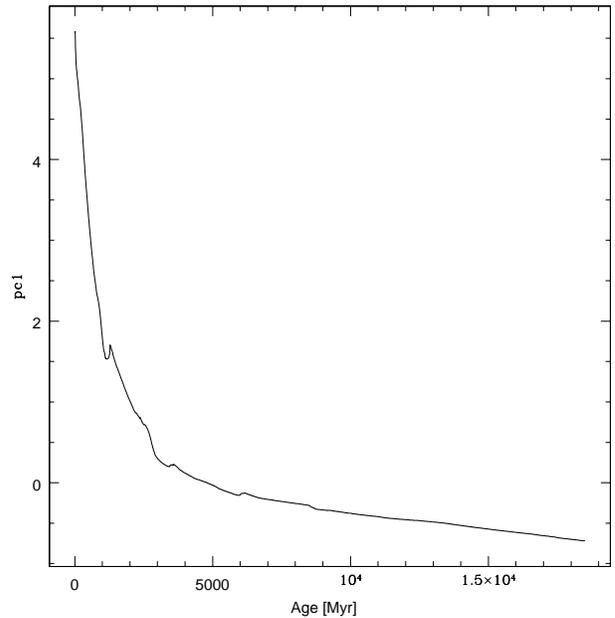,width=3.5in}
 \caption{{\bf Single burst:} the projection of the spectra on the first PC of figure~2
	versus age.}
\end{figure}

\begin{figure}
 \psfig{figure=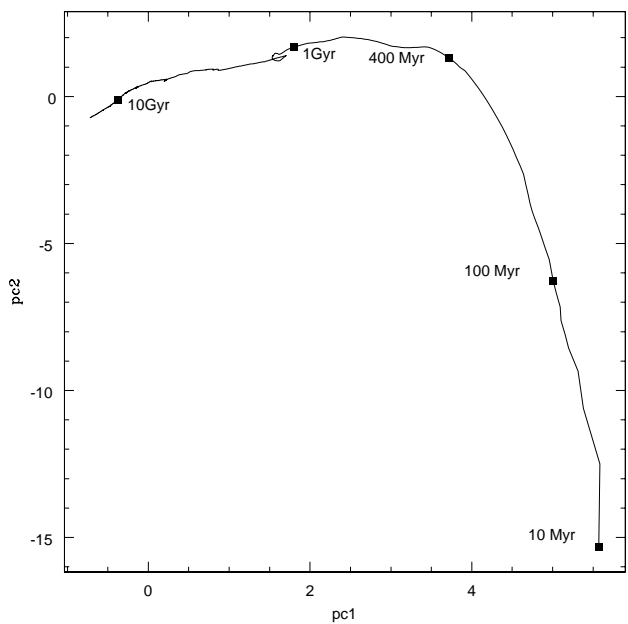,width=3.5in}
 \caption{{\bf Single burst:} the projection of the spectra on the first and second PCs of
          figure~2. The ages are marked.}
 
\end{figure}
\begin{figure}
 \psfig{figure=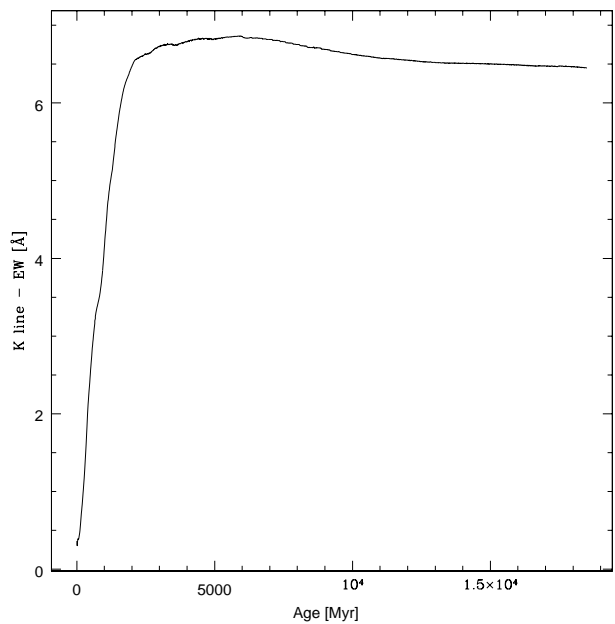,width=3.5in}
 \caption{{\bf Single burst:} The equivalent width of the Ca~K line ($\lambda$3933\AA) versus age. }
\end{figure}

We now turn to the analysis of very young populations with strong
emission lines, aged 0--15$\,$Myr. Figure~6 shows the mean spectrum and
the first PCs of a sample of galaxies with ages 0--14$\,$Myr and a
sampling interval of 1$\,$Myr. First note that the average spectrum is
indeed that of a very young galaxy: blue continuum with emission
lines.  The first PC contains mainly emission lines, with especially
strong H$\alpha$ and Oxygen lines. The energy source of this emission
is the UV photons (below the Lyman break at 912\AA) of very young blue
stars absorbed by the inter-stellar gas and then re-emitted by
recombination and radiative/collisional de-excitation.  The first PC
accounts for 99.5\% of the variation in this sample.  The second PC now
is basically a blue continuum, and is responsible for 0.7\% of the
variation. The third PC represents only 0.05\% of variation. It shows
absorption lines correlated with absorption below the Balmer break at
4000\AA, but also with persisting emission in H$\alpha$ and Oxygen
lines ---these being the strongest and the last ones to decay.

Figure~7 shows  the change of the PC1 and PC2 components with time. The
PC1 projection shows that emission lines dominate only at very young
ages after the initial star burst. Emission and/or partial filling of
absorption lines continue up to 6$\,$Myr, becoming much weaker at later
ages.  The PC2 projection shows that the continuum is maximally blue
about 3--5$\,$Myr after the burst, and then becomes less blue.
We interpret the sharp drop at 8 Million year as signalling the beginning of the death of the most
massive, bluest and most luminous stars.

\begin{figure}
 \psfig{figure=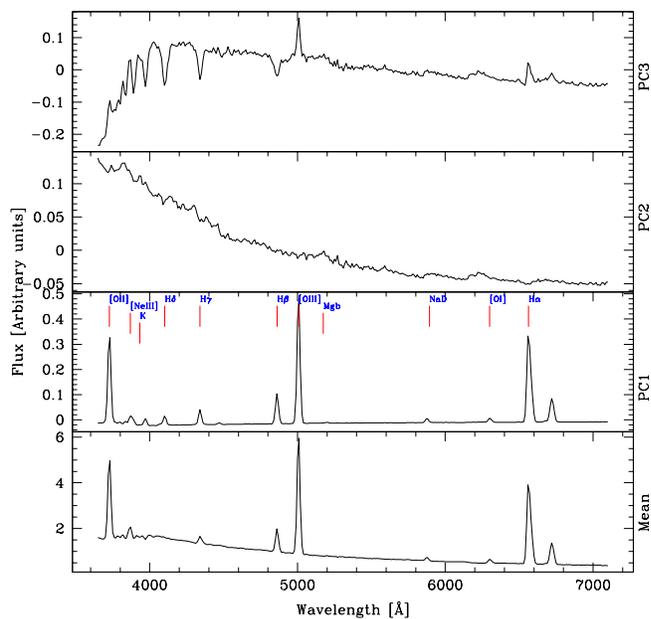,width=3.5in}
 \caption{{\bf Early single burst:} the mean of the synthetic spectra of galaxies formed in
a single instantaneous burst that takes place at $t=0$ in the age range
0--14$\,$Myr in equal time intervals of 1$\,$Myr (bottom). 
The first three PCs are plotted above it. PC1, PC2 and PC3 correspond to
a variance in the spectra of 99.5\%, 0.7\% and 0.05\%.}
\end{figure}
\begin{figure}
 \psfig{figure=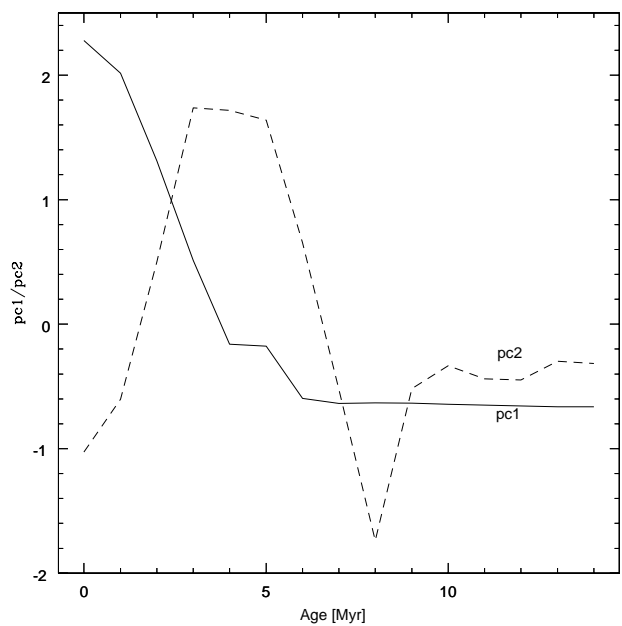,width=3.5in}
 \caption{{\bf Early single burst:} the projection of the 0--14$\,$Myr spectra on the first and 
second PCs of figure~6 versus age.}
\end{figure}

\section[] {Galaxies with a continuous star formation history}

In reality, the stellar populations of most galaxies are not formed in
a single instantaneous burst. 
Commonly continuous star-formation rate at time $t$ is modelled by 
\begin{equation}
SFR = (1/\tau)\exp(-t/\tau) \;,
\end{equation}
where $\tau$ is the $e$-folding time. 
Spiral galaxies seem to be best fitted by 
exponentially decreasing star formation rates (SFRs) with $e$-folding
times of the order of several Gyrs,  while ellipticals are best
fitted with models with much shorter star-formation time scales
($\tau\simeq 1Gyr$; see, e.g., Tinsley 1980; Bruzual 1983)

The difficulty of ascribing a certain age and star formation history to
a galaxy is that young age and a significant current star formation seem to
give the same effect on the spectrum:  blue light from young stars with strong
Balmer and other emission lines.  The reason for this is that the blue
young stars are the most luminous objects in the galaxy and dominate 
its spectrum.  One would like to find a way to separate these two important
parameters from the galaxy spectra. 

With this aim, we analyse a sample of galaxy spectra of ages
0.1--18.5$\,$Gyr with a time step of 100$\,$Myr and exponential SFR
with $e$-folding time $\tau$ in the range 0--7$\,$Gyrs.  
We have chosen the values
$\tau=0,0.1,0.5,1,2,3,5,7\,$Gyr ($\tau=0$ means single instantaneous
burst, as in our previous analysis).  Figure~8 shows the mean spectrum
and the first PCs of this sample.  The contributions of the PCs to the
variance are: PC1 97.08\%, PC2 2.75\%, PC3 0.13\%, and so again the
sample is fitted well by a line along PC1. Note the anti-correlation
between emission and absorption lines in it. This is in agreement with
our previous discussion of young single star-burst populations: the
competition between the physical processes of absorption in the stellar
atmospheres and gas emission produce stronger emission in the Oxygen
lines and H$\alpha$ and absorption in higher H lines. Still, at younger
ages we expect more emission, and as we shall see, this is shown by the
$negative$ values of PC2 (which has correlated $absorption$ lines).  The
prominent feature in PC3 is again absorption in Ca~K line. 
It is very interesting and encouraging that the first two PCs,
derived here from synthetic spectra, are very similar to the ones
obtained from a sample of observed galaxies ---see the analysis of
Folkes et al. (1996) using Kennicutt's (1992) data and the analysis
of the 2dF data (Folkes 1998).

\begin{figure}
 \psfig{figure=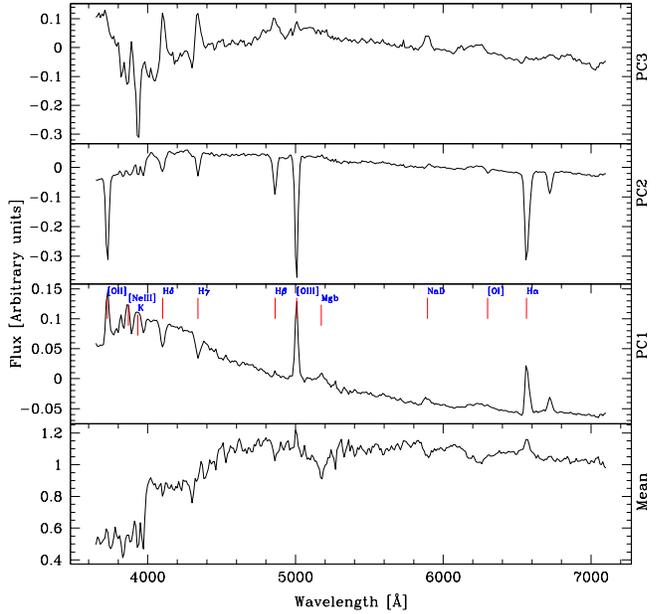,width=3.5in}
 \caption{{\bf Continuous star formation: } the mean of the synthetic spectra of galaxy models
with continuous exponentially decaying star formation histories aged 0.01--18.5$\,$Gyr
(bottom ---see text for details).  The first three PCs are plotted
above it.  PC1, PC2 and PC3 correspond to a variance in the spectra of 97.08\%, 
2.75\% and 0.13\%.} 
\end{figure}
\begin{figure}
 \psfig{figure=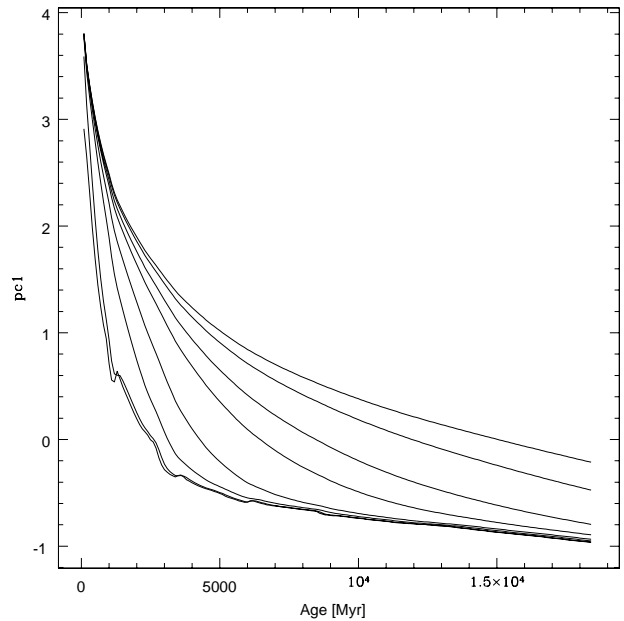,width=3.5in}
 \caption{{\bf Continuous star formation:} the projection of model galaxy spectra with
continuous star formation
histories on the first PC of figure~8 versus age. The lowest line 
corresponds to $\tau=0\,$Gyr, followed by $\tau=0.1,0.5,1,2,3,5,7\,$Gyr
in ascending order.
}	
\end{figure}

In figure~9 we plot the PC1 projection versus age for all the star
formation histories.  We see that, at a given age, galaxies with SFR
histories of larger $e$-folding times, (i.e., more star formation at
later ages), have bluer spectra and more emission lines.  This is to be
expected since when the star formation time scales become longer (larger
$e$-folding times), the ratio of the current star formation rate to to
average past SFR becomes larger (i.e., newly-formed stars contribute
more, in relative terms, than stars formed in the past ---recall that
we normalise the spectrum by total flux). Therefore, the ratio of the
current SFR over the average past SFR (or total stellar mass),
sometimes known as the `Scalo Parameter', is very important in
determining the spectrum of a galaxy.  A simple calculation shows that
this parameter is monotonically increasing with $\tau$ at any age for
exponentially-decaying star-formation histories.

The resulting ambiguity in age/SFR history is particularly severe for
redder galaxies, i.e. older ages.  To resolve it we  proceed to
look at the next PCs. Figure~10 is a projection on the PC1--PC2 plane.
First note the general trend: bluer (i.e. younger) galaxies with large
PC1 have enhanced emission lines (negative PC2).  With a starting age
of 100 Myr, the single burst spectra (top curve) already lack emission
lines, but show enhanced absorption lines instead.  This is in
accordance with our result in the previous section. The SFR histories
with $\tau$ between 0 and 2$\,$Gyrs are well separated around
PC1$\,=2.0$, indicating more emission strength for blue galaxies with
higher $\tau$. However looking back at Figure~9, this applies to
galaxies of $age<2\,$Gyr. For older and therefore redder galaxies with
smaller PC1 component, the curves in the PC1--PC2 plane tend to
converge, and do not provide a good separation.  Also for the range of
$\tau$ between 2 and 7$\,$Gyrs, all curves are very much the same over
the whole range.  We should therefore  turn to the third PC.

\begin{figure}
 \psfig{figure=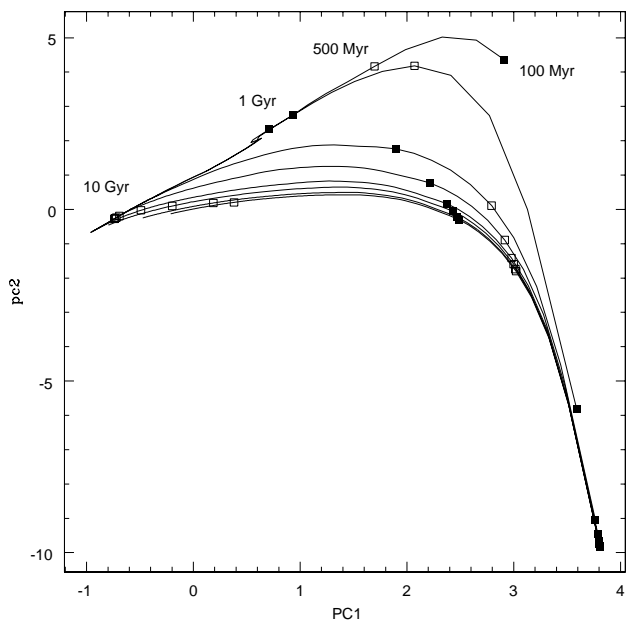,width=3.5in}
 \caption{{\bf Continuous star formation:} the projection of galaxy spectra with 
continuous star formation
histories on the first and second PCs of figure~8.  The top line
corresponds to $\tau=0$, followed by $\tau=0.1,0.5,1,2,3,5,7\,$Gyr in
descending order.  The alternating open and closed boxes indicate ages
of 0.1, 0.5, 1 and 10$\,$Gyr.
} 
\end{figure}

Figure~11 shows a projection on the PC1--PC3 plane. Although harder to
interpret, it does seem to give us some separation in the parameter range of
interest. A word of caution: the variance in PC3 (0.13\%) is
very small, and noise in real integrated spectra may cause a large
dispersion here. The extent of noise degradation is discussed in
section~7.

\begin{figure}
 \psfig{figure=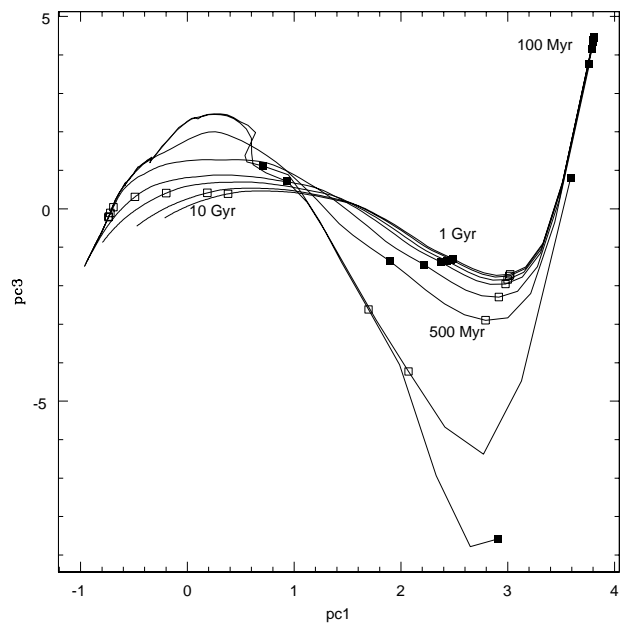,width=3.5in}
 \caption{{\bf Continuous star formation:} the projection of galaxy spectra with 
continuous star formation
histories on the first and third PCs of figure~8. At the left of the
figure, the top line corresponds to $\tau=0$, followed by
$\tau=0.1,0.5,1,2,3,5,7\,$Gyr in descending order. The alternating open
and closed boxes indicate ages of 0.1, 0.5, 1 and 10$\,$Gyr.
}
\end{figure}

\section[] {The effect of metallicity}

In this section address the issue of different metalicities in the
stellar populations of the galaxies and their impact on the spectra.
Again, we want to to separate the effects of metallicity and age on the
integrated galaxy spectra. Specifically, we apply it to old and simple
stellar populations, which could represent those of elliptical
galaxies.

Worthey (1994) investigated this problem by computing colours and
individual line indices in elliptical galaxy models, and studying the
variation of the different observables individually when changing age
and metallicity.  Here we apply the PCA technique to the problem,
studying the global properties of the spectra simultaneously, instead
of one feature at a time.  As described in section~2, we use the
Multi-metallicity Isochrone Synthesis Spectral Evolutionary Code of
Bruzual \& Charlot (1996), based on theoretical stellar spectra.  The
six different metalicities available have the following 
mass fraction $Z$  of elements heavier than
Helium: $Z=0.004,0.040,0.080,0.0200,0.0500$ and $0.1000$. In this
scale, the solar metallicity is about $Z=0.02$.  Using the BC models,
we sampled ages between 250 and 18500$\,$Myr with a step of 250 Myr,
and performed PCA on a single burst model sample including all six
metalicities. Emission lines become negligible in this age range for
single burst models.

Figure~12 shows the mean spectrum and the first PCs of this sample.
The contributions of the PCs to the variance are: PC1 96.41\%, PC2
2.03\%, PC3 0.34\%.  The average spectrum and the first PC are very
similar to those shown in figure~2 for a single solar metallicity
population, although now there is a larger variance associated to the
metallic lines, as expected for a sample of galaxies which contain a
mixture of metalicities. In PC1 some of the metallic lines appear as
`emission' features (e.g., the NaD, Mgb, and Fe lines).   These are in
fact absorption lines, but anti-correlated with Balmer absorption lines
and blue colours. This in accordance with the results of Worthey
(1994).  PC2 is very different from the one shown in figure~2.  The
variance in it is bigger than in the case of single solar metallicity,
reflecting the larger variance in the sample due to metallicity effects.
It contains metallic absorption features:  NaD, Mgb, Fe5015, Fe5782.
The most dominant are NaD\footnote {Nad is a resonant line, and could be affected
by interstellar absorption, which the models do not include.} and Mgb absorption lines.
PC2 also contains
Balmer absorption lines, this time positively correlated with the
metallic lines. We explain this in the discussion of figure~14 below.

\begin{figure}
 \psfig{figure=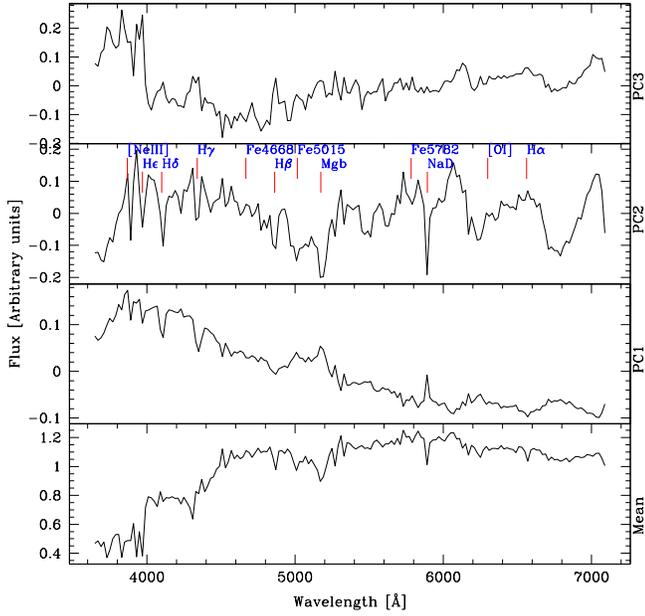,width=3.5in}
 \caption{{\bf Metaliciy effects:} the mean of the synthetic spectra of stellar populations 
with different metalicities aged 100--1850$\,$Myr (bottom), and  
the first three PCs plotted above it. PC1, PC2 and PC3 correspond to a variance in 
the spectra of 96.41\%, 2.03\% and 0.34\%}
\end{figure}

In figure~13 we plot the PC1 projection versus age for all the
metalicities.  Since PC1 as a blue continuum slope, we see a clear
uniform tendency for galaxies to become redder with increasing
metallicity.  The exception are the models with the highest metallicity,
which become much bluer after 14Gyr. Bruzual \& Charlot (1996)
explanation for that behaviour is the appearance of AGB-manqu\'e stars
at $Z=0.1$, which due to a very efficient mass loss skip the AGB phase
and instead go through a long lived hot HB phase.  That could explain
the UV-upturn seen in the spectra of massive (i.e., high metallicity)
ellipticals.  However, these authors warn that this particular result
should be taken with caution, as the evolutionary tracks of high
metallicity stars are quite uncertain. At any rate, $Z=0.1$ is 5 times
the solar metallicity and observationally very uncommon.

\begin{figure}
 \psfig{figure=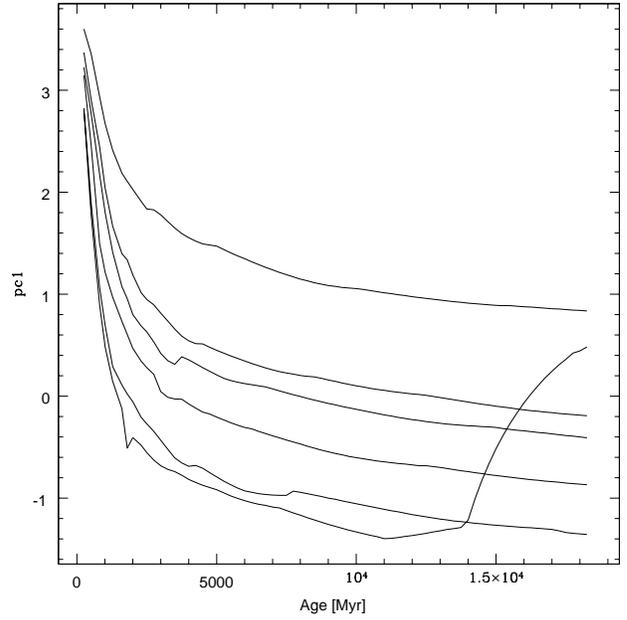,width=3.5in}
 \caption{{\bf Metallicity effects:} the projection of spectra of different metallicity on the
first PC of figure~12 versus age. The bottom line corresponds to $Z=0.1$, followed by
$Z=0.05,0.02,0.008,0.004$ and $0.0004$}
\end{figure}

In figure~14 we plot the projection of the sample on the PC1--PC2
plane.  In general, a population with higher metallicity occupies a
locus with higher PC2 for a given PC1 (or colour). However, this is not
true for a given age: for 10$\,$Gyr, higher metallicity gives higher
PC2, but for 1$\,$Gyr it gives lower PC2. The explanation for this as
follows: in general, for a given age, the strength of the Balmer
absorption lines decreases with metallicity (as already seen by Worthey
1994).  At 1$\,$Gyr the Balmer lines are strong and dominate the
projection on PC2. At 10$\,$Gyr, the Balmer lines are weak and the
metallic absorption lines dominate the projection on PC2.  The
correlation between metallic and Balmer lines in PC2 indicates that
this is indeed the case, in a statistical sense, for a given PC1 (or
colour).

We conclude that the fact that the first PC already accounts for
96.41\% of the variation shows that there is indeed an age-metallicity
degeneracy, but the use of the second PC allows us to separate them in
principle. 
It may be argued that, from its definition, the PCA method
is  the optimal way to achieve such a separation (at least using linear
methods). PCA is also simple, since there is no need to calculate a
multitude of photometric colours and spectral indices: all the spectral
features are taken into account {\it simultaneously, in an optimal way}.

\begin{figure}
 \psfig{figure=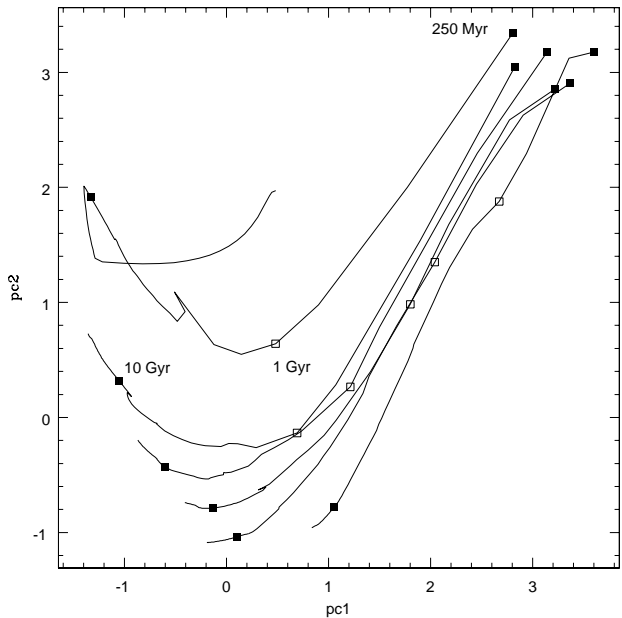,width=3.5in}
 \caption{{\bf Metallicity effects:} 
The projection of spectra of different metallicity on the PC1--PC2
plane. The top line corresponds to $Z=0.1$ followed by
$Z=0.05,0.02,0.008,0.004,0.0004$ (the lines do not cross over in the
middle).  The alternating open and closed boxes indicate ages of $0.25,
1$ and $10\,$Gyr.}
\end{figure}

\section[] {The effect of noise} 

When it comes to the analysing of real observed spectra, one of the major
concerns is the signal to noise ($S/N$) ratio.  In previous sections we
have shown that information contained in the second or third PCs may
enable us to separate the effects on the spectra of different star
formation histories or metalicities from age, thus determining all
these parameters simultaneously. However, as we have shown, the
relative variance contained in the second and third PCs is generally
very small, of order of one percent. Therefore, in order to suggest PCA
as a practical method for the above task,  we need to pay careful
attention to the effect of noise. 
Noise affects both the eigenvectors and the projection of objects into the 
PCA space.

Let $N_p$ be the total number of photon counts in a given spectral
bin, and let us assume that photon number counts obeys Poisson
statistics.  Usually $N_p$ will get contributions both from the galaxy
flux and from the sky spectrum, however for the sake of our analysis
we ignore the sky contribution (See Folkes et al. 1996 for simulations
including sky contribution).
Define the signal $S$ to be the
expectation value of the number of counts: 
\begin{equation}
S=\langle N_p \rangle\;.
\end{equation}
Then the signal to noise ratio in this bin is just:
\begin{equation}
S/N=\sqrt {S}\;.
\end{equation}
The noise variance is added to the diagonal of the covariance matrix
(eq. 3), as the noise in different bins is supposed to be uncorrelated. 
For a realistic galaxy spectrum the 
noise in each bin is different (as it depends on the galaxy flux at each bin).
Hence the addition of noise may rotate the eigenvectors to different
directions compared with the noise-free case (if the noise is constant per
bin then the direction of the eigenvectors is not affected, only the eigenvalues
are changed). However,  one expects, and our experiments confirm, that the first 
eigenvectors corresponding to large intrinsic 
variance in the clean data will be little affected by noise.
Also note that the noise in the eigen-spectra will be reduced with the size of
the sample spectra, because of the averaging of the covariance matrix
over all the spectra in the sample.

Another effect of the noise in spectra will be that, even if
eigenvectors do not chance direction, noise in a given spectrum will
change its projection into the PCA space.  This will introduce scatter
in projection plots, such as pc1-pc2 plots presented in this work. It
will increase the PCA eigenvalue corresponding to any PC. We wish to
derive here a simple mathematical expression for this effect, under the 
following assumptions.  Let us consider a simplified case
in which the clean spectrum is a flat line. Then the $S/N$ is constant
over the whole wavelength (even if the clean spectrum is not flat,
then, to a first approximation, we can take the $S/N$ to be a constant,
equal to the average signal to noise over the observed wavelength
range).  Secondly, let us examine a sample of galaxies with the same
apparent magnitude, so that the $S/N$ is the same for all galaxies in the
sample. Also let us assume that the eigenvectors of 
interest are largely unaffected by the noise (but see below
for a more careful examination of this assumption).

Then we can calculate the noise in the projection coefficients and the
resulting change in eigenvalues as follows. As usual, the spectra are
first normalized to have a constant total flux.  For convenience
we choose a normalization such that the average normalized counts per bin
is 1.  I.e, we are dividing $N_p$ by S. Then the variance of $N_p$ in
this normalization, which we denote by $V$, is:
\begin{equation}
V=Var(N_p/S)=Var(N_p)/S^2=1/S=1/(S/N)^2.
\end{equation}
Therefore $V$ is the (renormalized) variance of noise in each bin. Under
our assumptions this is the same for all bins, i.e, the noise is
isotropic. Hence, after the orthogonal transformation to the PCA
basis, the contribution of noise to the variance in each PC direction
will be the same for all PCs, and equal to $V$.  In other words, when
calculating the projection coefficients of a given galaxy spectra on a
given PC, each coefficient will include noise with a variance of $V$.

If in the absence of noise the eigenvalue of a given 
eigenvector $i$ of the covariance matrix is $\lambda_i$, then in the
presence of noise this eigenvalue becomes $\lambda_i+V$. 
Taking a square root to 
move from variances to standard deviations, we conclude that the ratio
\begin{equation}
Q_i\equiv \sqrt{V\over\lambda_i}={1\over(S/N)\cdot\sqrt{\lambda_i}}
\end{equation}
gives us a measure of the relative spread in PCA projection 
coefficients due to noise, as compared
with the internal variance in the data due 
to galaxies having different spectra. The requirement for
a projection on a given PC to 
be relatively noise free is therefore $Q_i\gg 1$, or:
\begin{equation}
S/N\gg 1/{\sqrt\lambda_i} \;.
\end{equation}
This last expression, although a crude approximation, 
is useful for planning observations 
to discriminate between 
age/sfr/metallicity using the PC projections presented above.   

\begin{figure}
 \psfig{figure=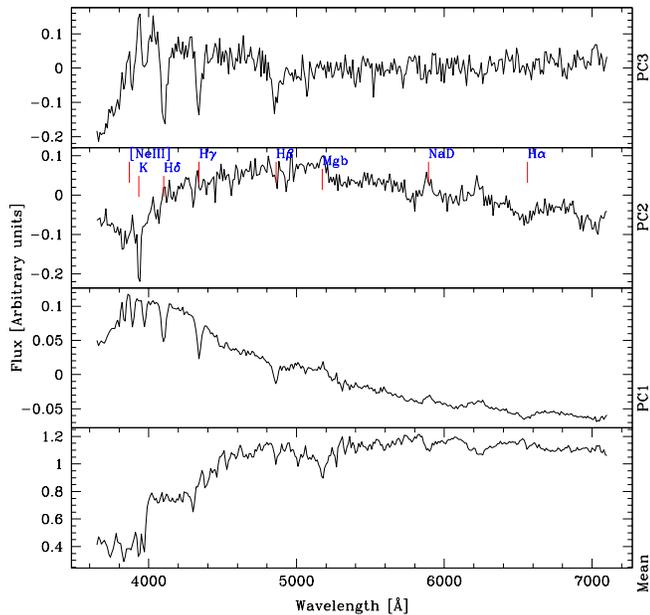,width=3.5in} 
 \caption{{\bf The effect of noise,} with a $S/N=10$, on the PCs 
 of single burst stellar population spectra, to be compared with figure~2.
  PC1, PC2 and PC3 correspond to a variance in the spectra of 52.2\%, 0.6\% and 0.4\%}
\end{figure}

\begin{figure}
 \psfig{figure=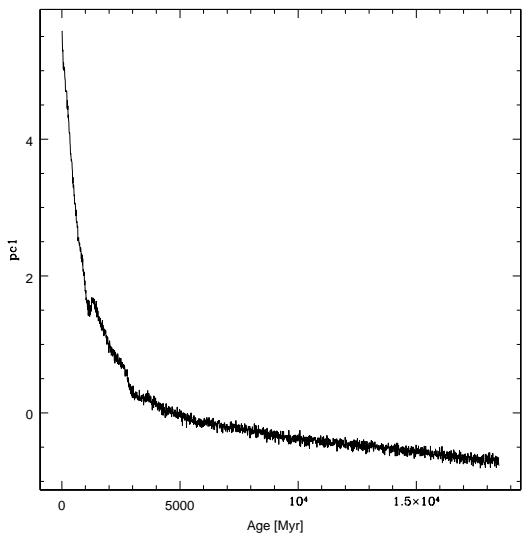,width=3in}
 \caption{{\bf Effect of noise:} The projection of the noisy spectra on the first PC of figure~15
	versus age, to be compared with figure~3.}
\end{figure}

\begin{figure}
 \psfig{figure=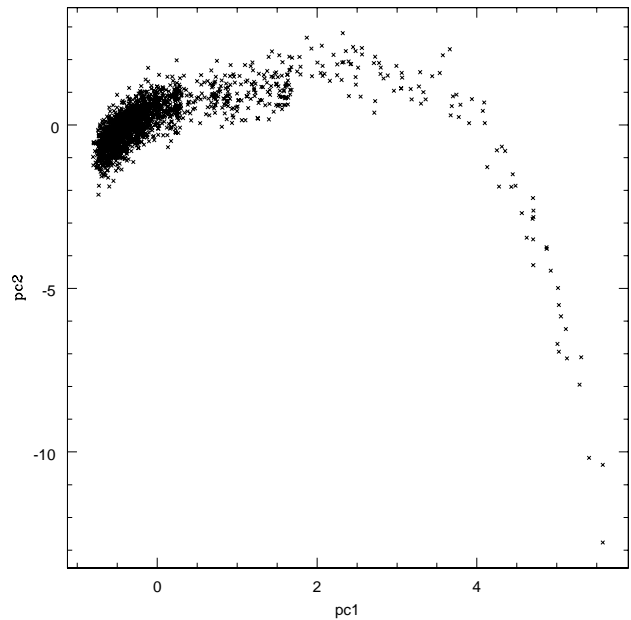,width=3.5in}
 \caption{{\bf Effect of noise:} The projection of the noisy spectra on the first and second PCs of
          figure~15, to be compared with figure~4.}
 
\end{figure}

For example, in the study of
single burst galaxies we found that $\lambda_1=3.96$, $\lambda_2=0.035$ and
$\lambda_3=0.020$. 
Therefore, for PC1, PC2 and PC3 projections to be meaningful 
the signal-to-noise $S/N$ 
must be larger than 0.5, 5.3 and 7.1 respectively
(per $10\AA$  spectral bin which is the resolution considered here).  
The PC1 projection is therefore negligibly affected
by noise while PC2 and PC3 
require very high $S/N$.

In the analysis of continuous star formation histories there is a
larger variance in PC2, corresponding to (i.e., requiring) a minimum
$S/N=2$. For PC3, a $S/N=10$ is required.  In the analysis of galaxies
with different metalicities we get $S/N=3.6$ for PC2 and $S/N=5.3$ for
PC3.  For example, 
to really differentiate between two consecutive metallicity
curves of figure~14, 
a significantly smaller scatter then the whole pc2 variance is
required, and the required $S/N$ should be scaled up accordingly.

We concluded in previous sections that to separate age from SFR
history, one will need PC3, while to separate age from metallicity PC2
is enough. Therefore, the above numbers indicate that a $S/N > 10$ per
wavelength element (at $\simeq10$\AA\ resolution) may already enable
some metallicity discrimination 
(between the highest and lowest metallicity values), 
but higher quality spectra are required to
determine the star formation history separately from the age.

To illustrate in a more realistic way the effect of noise discussed
above, we have performed a numerical simulation.  We returned to the
analysis presented in section 4 of single burst population with ages
0--18.5$\,$Gyr, but this time with (a Poisson) noise added to the
synthetic spectra. For each galaxy, we assumed a magnitude such that
the average photon counts per bin is 100. This corresponds to a $S/N=10$
in our simplified analysis above. However, in the simulation we did
include a $S/N$ which properly changes  from bin to bin according to flux. 
Figure~15 shows the resulting PCs, to be compared with figure~2. The noise little affects 
the average spectrum and the first PC, while the third PC is more
significantly degraded. Also, PC1, which previously accounted for
98.5\% of the variance of the sample, now accounts for only 52.2\%
(PC2 0.6\%, PC3 0.4\%), and much of the variance, due to noise, is
more or less evenly distributed between PC's higher than the third
one.  Figure~16 shows the projection of the noisy spectra on the first
PC, to be compared with figure~3. As expected, the PC1 projection is
affected by the noise only slightly.  Figure~17 shows the projection
of the noisy spectra on the PC1--PC2 plane, to be compared with
figure~4.  Significant noise enters the PC2 projection, but it is
still less than the total variance in PC2, as could be expected from
the simple calculations above.

\section[] {Discussion}

In this study we applied PCA  to synthetic spectra with
different ages, star formation histories and metalicities.
The PCA provides a more objective
and informative alternative to diagnostics by individual spectral
lines,  as in the PCs represent all the spectral
features {\it simultaneously}.
In particular it is encouraging that our first two synthetic PCs
agree with a similar analysis on observed spectra obtained by
Kennicutt (e.g. Sodre \& Cuevas 1997; Folkes et al. 1996)  
and the new 2dF redshift survey (Folkes 1998).

Our first experiments with single burst populations show, as expected,
the significant effect of the sampling on the resulting PCs.  I.e,
when one samples young galaxies the first PCs reflect the main
variance in that young age which is different from that for an older
population. In examining data from big spectral surveys we do not know
a priori how many galaxies are of each age.
Therefore it seems best to derive the PCs (eigenvectors)
from the observational data, and then project on them both the
observational data and the synthetic model spectra in order to find
the best fit parameters (age, SFR, metallicity) for the observed
galaxies. The disadvantage of deriving the PCs from the observational
data is the effect of noise on the eigenvectors themselves.  However,
any systematic effect due to noise anisotropy will be small for the
first eigenvectors, and the random effect will be reduced with the 
inverse square root of the 
number of spectra, which is encouraging for
the coming big spectral surveys (2dF, Sloan). Alternatively, one can
project the observational data on the theoretical PCs derived from
some pre-chosen synthetic sample of ages (and other parameters). The
statistical composition the synthetic spectra may be chosen on some
theoretical grounds, or experimentally, from the first procedure just
outlined. We shall discuss elsewhere various approaches to this
`inverse problem'.
 
When there are only two parameters such as age and SFR or age and
metallicity, our experiments suggest the possibility to determine them
separately from the first two or three PCs, given a good enough $S/N$
($S/N\gg 10$ for metallicity and better for SFR discrimination).  In
practice all three are involved, including others, such as chemical
evolution, extinction by dust, etc.  However assuming the first three
are responsible for the major variance in the spectra, our analysis
suggests that a parameter fit using the PCA method may be successful
for high quality spectra. This analysis highlights the need for a large
sample of high signal-to-noise of galaxy spectra.

\section*{Acknowledgements}

We thank S. Folkes and K. Glazebrook for helpful discussions and 
G. Bruzual, S. Charlot, M. Fioc \& B. Rocca-Volmerange
for providing easy access to their model calculations.  
S. Ronen is grateful for a scholarship from Trinity College, Cambridge,
and A. Arag\'on-Salamanca acknowledges generous financial support by the
Royal Society.

\bsp

\label{lastpage}

\end{document}